\documentclass[conference]{IEEEtran}

\usepackage{bm}
\usepackage{algorithm}
\usepackage{algorithmic}
\usepackage{amsmath}
\usepackage{amssymb}
\usepackage{amsfonts}
\usepackage{graphicx}
\usepackage{epsfig}
\usepackage{subfigure}
\usepackage{psfrag}
\usepackage{cite}
\usepackage{latexsym}
\usepackage{url}
\usepackage{color}
\usepackage{multirow}
\IEEEoverridecommandlockouts

\PassOptionsToPackage{bookmarks={false}}{hyperref}

\begin{document}
\title{UAV-Enabled Wireless Power Transfer with Directional Antenna: A Two-User Case
\thanks{J. Xu is the corresponding author.}}
\author{Yundi Wu$^1$, Jie Xu$^2$, and Ling Qiu$^1$\\
$^1$School of Information Science and Technology, University of Science and Technology of China\\
$^2$School of Information Engineering, Guangdong University of Technology\\
E-mail:~wyd57@mail.ustc.edu.cn,~jiexu@gdut.edu.cn,~lqiu@ustc.edu.cn
}
\IEEEspecialpapernotice{(Invited Paper)}

\maketitle

\begin{abstract}
This paper considers an unmanned aerial vehicle (UAV)-enabled wireless power transfer (WPT) system, in which a UAV equipped with a directional antenna is dispatched to deliver wireless energy to charge two energy receivers (ERs) on the ground. Under this setup, we maximize the common (or minimum) energy received by the two ERs over a particular finite charging period, by jointly optimizing the altitude, trajectory, and transmit beamwidth of the UAV, subject to the UAV's maximum speed constraints, as well as the maximum/minimum altitude and beamwidth constraints. However, the common energy maximization is a non-convex optimization problem that is generally difficult to be solved optimally. To tackle this problem, we first ignore the maximum UAV speed constraints and solve the relaxed problem optimally. The optimal solution to the relaxed problem reveals that the UAV should hover above two symmetric locations during the whole charging period, with the corresponding altitude and beamwidth optimized. Next, we study the original problem with the maximum UAV speed constraints considered, for which a heuristic hover-fly-hover trajectory design is proposed based on the optimal symmetric-location-hovering solution to the relaxed problem. Numerical results validate that thanks to the employment of directional antenna with adaptive beamwidth and altitude control, our proposed design significantly improves the common energy received by the two ERs, as compared to other benchmark schemes.
\end{abstract}

\newtheorem{definition}{\underline{Definition}}[section]
\newtheorem{fact}{Fact}
\newtheorem{assumption}{Assumption}
\newtheorem{theorem}{\underline{Theorem}}[section]
\newtheorem{lemma}{\underline{Lemma}}[section]
\newtheorem{corollary}{\underline{Corollary}}[section]
\newtheorem{proposition}{\underline{Proposition}}[section]
\newtheorem{example}{\underline{Example}}[section]
\newtheorem{remark}{\underline{Remark}}[section]

\newcommand{\mv}[1]{\mbox{\boldmath{$ #1 $}}}
\setlength\abovedisplayskip{6pt}
\setlength\belowdisplayskip{6pt}

\section{Introduction}
Radio frequency (RF) signals-enabled wireless power transfer (WPT) has been recognized as a promising technique to provide convenient and sustainable wireless energy supply for energy-constrained electronic devices \cite{LuWang2015}. As a result, WPT has enabled abundant applications in wireless networks, such as simultaneous wireless information and power transfer (SWIPT) \cite{ClerckxZhang}, wireless powered communication network (WPCN) \cite{BiHoZhang2015}, and wireless powered mobile edge computing \cite{WangXuWang2017}. In conventional WPT systems, energy transmitters (ETs) are deployed at fixed locations to broadcast wireless energy towards distributed energy receivers (ERs). Due to the RF signal propagation over distances, such systems suffer from low end-to-end energy transfer efficiency, especially when ERs are far away from the ET.

With recent advancements in unmanned aerial vehicles (UAVs) \cite{ZengZhangLim2016}, a new WPT architecture, namely the UAV-enabled WPT,  has been proposed in \cite{XuZengZhang20171,XuZengZhang20172} to help improve the end-to-end energy transfer efficiency by employing UAVs as mobile ETs. Different from conventional ETs that are fixed, the mobile ETs in UAV-enabled WPT systems can fully exploit the UAVs' controllable mobility in three-dimensional (3D) airspace to adaptively adjust their locations over time (a.k.a. trajectory), so as to shorten the distances with intended ERs and accordingly improve the energy transfer efficiency. More specifically, \cite{XuZengZhang20171} considered the scenario with one UAV serving two ERs on the ground, in which the UAV adaptively controls its trajectory to balance the received energy tradeoff between the two ERs. \cite{XuZengZhang20172} considered a more general scenario with more than two ERs, in which the UAV aims to maximize the sum or the minimum of the received energy for these ERs via trajectory control. Furthermore, \cite{XieXuZhang} extended \cite{XuZengZhang20171,XuZengZhang20172} to a UAV-enabled WPCN, in which the moving UAV sends wireless energy to charge users on the ground, and the users use the harvested energy to send information (e.g., collected environmental data) back to the UAV. Despite the research progress, the above works assumed that the UAV is equipped with an omni-directional antenna that radiates RF signals in all directions, and the UAV flies at a fixed altitude over time. This, however, may not be efficient for air-to-ground (A2G) energy broadcasting, where ERs on the ground are located at given directions below the UAV in the sky. Therefore, it is expected that directional antenna (see, e.g., \cite{HeZhangZengZhang2018,Balanis:Book}) could be a viable means to further improve the energy transfer performance of the UAV-enabled WPT system, thus motivating our investigation in this paper.

For the purpose of exposition, this paper considers that a UAV is dispatched to broadcast wireless energy to two ERs on the ground, and the UAV has a directional antenna with beamwidth adjustable over time for increasing the energy transfer efficiency. Despite the benefit, however, the UAV-enabled WPT system with directional antenna faces various design challenges in the UAV's trajectory, altitude, and beamwidth control to achieve the optimal energy transfer performance. First, the received energy at the two ERs critically depends on the UAV's locations over time (or trajectory). If the UAV moves closer to one ER, then the received power at this ER will increase, but that at the other ER will decrease. This thus results in an energy trade-off between the two ERs in the UAV's trajectory design. Next, there is another tradeoff in controlling the UAV's flying altitude and beamwidth. On one hand, in order to charge each ER most efficiently, the UAV should narrow down the beamwidth and decrease the altitude to maximize the antenna gain and minimize the signal propagation pathloss, respectively. On the other hand, to broadcast energy towards the two ERs, the UAV needs to stay above a certain altitude and widen the beamwidth for ensuring that the transmitted signal beam covers both ERs at the same time. By combining the above tradeoffs, how to jointly design the trajectory, altitude, and beamwidth of the UAV for fairly maximizing the two ERs' received energy is a crucial but challenging task to be tackled.

Specifically, we are interested in maximizing the common (or minimum) energy received at the two ERs over a particular finite charging period, by jointly optimizing the UAV's trajectory, altitude, and beamwidth, subject to the UAV's maximum speed constraints, as well as the maximum/minimum altitude and beamwidth constraints. However, the common energy maximization is a non-convex optimization problem that is difficult to be solved optimally. To tackle this problem, we first consider a relaxed problem by ignoring the maximum UAV speed constraints and solve it optimally. The optimal solution to the relaxed problem reveals that the UAV should hover above two symmetric locations during each half of the whole charging period, with the corresponding altitude and beamwidth optimized. Next, for the original problem with the maximum UAV speed constraints considered, we propose a heuristic hover-fly-hover trajectory design based on the optimal solution to the relaxed problem. Finally, numerical results validate that our proposed design with directional antenna significantly improves the common energy received by the two ERs, as compared to the conventional UAV-enabled WPT system with omni-directional antenna and the benchmark scheme when the UAV statically hovers at an optimized location.

\section{System Model}
We consider a two-user UAV-enabled WPT system, where a UAV is dispatched to deliver wireless energy to charge $K=2$ ERs on the ground. Let $\mathcal{K} \triangleq \{ 1,2 \}$ denote the set of ERs. We suppose that each ER $k \in \mathcal{K}$ has a fixed location, denoted by $(\hat{x}_k,0,0)$ in 3D space, where $\hat{x}_1=-\frac{D}{2}$ and $\hat{x}_2=\frac{D}{2}$, with $D>0$ denoting the distance between the two ERs. We consider a finite charging period $\mathcal{T} \triangleq (0,T]$, with duration $T>0$. During this period, the UAV can change its location in 3D with the time-varying coordinate denoted as $(x(t),y(t),H(t)), t \in \mathcal{T}$. Without loss of optimality, we consider that $y(t)=0,\forall t \in \mathcal{T}$, such that the UAV always stays above the line between the two ERs for efficient WPT. Let $H_{\text{min}}$ and $H_{\text{max}}$ in meter (m) denote the minimum and maximum altitudes of the UAV, respectively. We then have the UAV altitude constraints as
\vspace{-0.5em}
\begin{align}\label{altitude cons}
H_{\text{min}} \le H(t) \le H_{\text{max}}, \forall t \in \mathcal{T}.
\end{align}
Furthermore, let $V>0$ in meter/second (m/s) denote the maximum UAV speed. It then follows that
\vspace{-0.5em}
\begin{align}\label{speed cons}
\sqrt {\dot{x} ^2 (t)+\dot{H} ^2 (t)} \le V,\forall t \in \mathcal{T},
\end{align}
where $\dot{x}(t)$ and $\dot{H}(t)$ denote the time-derivatives of $x(t)$ and $H(t)$, respectively.

We assume that the UAV is equipped with a directional antenna whose beamwidth is adjustable over time. Similarly as in \cite{HeZhangZengZhang2018}, it is assumed that the azimuth and elevation half-power beamwidths of the UAV antenna are equal, which are both denoted as $2\Theta(t)$ in radians (rad) at time instant $t \in \mathcal{T}$. Let $0 <\Theta_{\text{min}} \le \Theta_{\text{max}} \le \pi/2$ denote the minimum and maximum beamwidths of the directional antenna, and the we have
\vspace{-0.5em}
\begin{align}\label{beamwidth cons}
\Theta_{\text{min}} \le \Theta(t) \le \Theta_{\text{max}},\forall t \in \mathcal{T}.
\end{align}
For ease of presentation and to avoid trivial solutions, in this paper we consider the scenario when $\tan \Theta_{\text{max}} \ge D/(2H_{\text{max}})$ and $\tan \Theta_{\text{min}} \le D/H_{\text{min}}$ hold, such that when the maximum altitude and beamwidth are adopted, the UAV¡¯s transmitted signal beam can cover the two ERs simultaneously; but when the minimum altitude and beamwidth are adopted, the UAV's transmitted signal beam cannot cover the two ERs at the same time\footnote{Notice that when $\tan \Theta_{\text{max}} < D/(2H_{\text{max}})$, the UAV can send energy to only one ER at a time. In this case, the UAV should charge the two ERs one by one individually, thus making the design problem trivial to solve. On the other hand, when $\tan \Theta_{\text{min}} > D/H_{\text{min}}$, the UAV's transmitted signal beam can always cover the two ERs, and therefore, the UAV should fly at the minimum altitude and uses the minimum beamwidth to efficiently charge the two ERs at the same time. In this case, the common-energy minimization problem in (P1) becomes identical to that with omni-directional antenna in \cite{XuZengZhang20172}, where only the UAV trajectory needs to be designed.}. According to Eq. (2-51) in \cite{Balanis:Book} and Eq. (1) in \cite{HeZhangZengZhang2018}, the antenna gain in direction $(\hat{\theta},\hat{\psi})$ is given by
\vspace{-0.5em}
\begin{align}
&G(\hat{\theta},\hat{\psi})=\nonumber\\
&\begin{cases} G_0/\Theta^2,&\text{if}~-\Theta(t) \le \hat{\theta} \le \Theta(t),-\Theta(t) \le \hat{\psi} \le \Theta(t)\\0,~&\text{otherwise}, \end{cases}
\end{align}
where $G_0 =\frac{30000(\frac{\pi}{180})^2}{2^2} \approx 2.2846$. Notice that if the UAV has the coordinate $(x(t),0,H(t))$ at time instant $t \in \mathcal{T}$, then ER $k \in \mathcal{K}$ is located at direction $(\theta_k(t),\psi_k(t))$ with antenna gain $G(\theta_k(t),\psi_k(t))$, where $\theta_k(t)=\arctan((x(t)-\hat{x}_k)/H(t))$ and $\psi_k(t)=0$, i.e.,
\vspace{-0.5em}
\begin{align}
G(\theta_k(t),\psi_k(t))=&\begin{cases} \frac{G_0}{\Theta^2(t)},&~\text{if}~\tan\Theta(t) \ge \frac{|x(t)-\hat{x}_k|}{H(t)}\\0,&~\text{otherwise}. \end{cases}
\end{align}

In practice, the wireless channel from the UAV to each ER is dominated by the LoS component. Therefore, as commonly adopted in prior works \cite{XuZengZhang20171,XuZengZhang20172}, we consider the free-space path-loss model, in which the channel power gain from the UAV to ER $k \in \mathcal{K}$ is denoted as $h_k(t) =\beta _0 d_k^{ - 2}(t)$. Here, $d_k(t)=\sqrt {(x(t)-\hat{x}_k)^2+H^2(t)} $ denotes their distance at time instant $t \in \mathcal{T}$ and $\beta _0$ denotes the channel power gain at a reference distance of $d_0=1~\mathrm{m}$. Furthermore, we assume that the two ERs are each equipped with one omni-directional antenna with unit gain. Assuming that the UAV adopts a constant transmit power $P$, the received power by ER $k$ at time instant $t$ is
\vspace{-0.5em}
\begin{align}
&Q_k(x(t),H(t),\Theta(t))=Ph_k(t)G(\theta_k(t),\psi_k(t))\nonumber\\
=&\begin{cases} \frac{\beta_0G_0P}{\Theta^2(t)((x(t)-\hat{x}_k)^2+H^2(t))},&~\text{if}~\tan\Theta(t) \ge \frac{|x(t)-\hat{x}_k|}{H(t)}\\0,&~\text{otherwise}. \end{cases}
\end{align}
Accordingly, the total RF energy received by ER $k \in \mathcal{K}$ over the whole charging period is expressed as
\vspace{-0.5em}
\begin{align}
E_k(\{x(t),H(t),\Theta(t)\})=\int_0^T Q_k(x(t),H(t),\Theta(t)) \text{d}t.
\end{align}
Notice that at each ER, the received RF signals are converted into direct current (DC) signals via a rectifier to charge the rechargeable battery. In practice, the harvested DC power is a nonlinear function with respect to the RF power, and this function critically depends on various issues such as the input RF power level, the structure of energy harvesting circuits, and the transmit signal waveform \cite{ClerckxZhang,Clerckx2016,Boshkovska2015}. For ease of exposition, we only focus on the received RF energy in this work.

In order to fairly transfer energy to the two ERs, we are interested in maximizing the common or minimum energy received by the two ERs (i.e., $\min_{k \in \mathcal{K}}E_k(\{x(t),H(t),\Theta(t)\}$) over the duration $T$, by jointly optimizing the half-power beamwidth $\{\Theta(t)\}$, altitude $\{H(t)\}$, and trajectory $\{x(t)\}$ of the UAV over time, subject to the altitude constraints in \eqref{altitude cons}, the maximum speed constraints in \eqref{speed cons}, and the beamwidth constraints in \eqref{beamwidth cons}. Mathematically, the
optimization problem is formulated as
\vspace{-0.5em}
\begin{align}
\text{(P1)}:&\max_{\{x(t),H(t),\Theta(t)\}}~\min_{k \in \mathcal{K}}~ \int_{0}^T {Q}_k(x(t),H(t),\Theta(t)) \mathrm{d} t \nonumber\\
&~~~~~~~~~\mathrm{s.t.}~~~~~~~
 \eqref{altitude cons},~\eqref{speed cons},~\text{and}~\eqref{beamwidth cons}.\nonumber
\end{align}
Note that in (P1), there are infinite number of optimization variables (i.e. $\{x(t)\}$, $\{H(t)\}$ and $\{\Theta(t)\}$) that are continuous over time, and the objective function is non-concave in general. Therefore, (P1) is a non-convex optimization problem that is difficult to be solved optimally. To tackle this problem, in Section III we first optimally solve a relaxed problem of (P1) (i.e., (P2) in the following), by ignoring the maximum UAV speed constraints in \eqref{speed cons}.
\vspace{-0.5em}
\begin{align}
\text{(P2)}:&\max_{\{x(t),H(t),\Theta(t)\}}~\min_{k \in \mathcal{K}}~ \int_{0}^T {Q}_k(x(t),H(t),\Theta(t)) \mathrm{d} t \nonumber\\
&~~~~~~~~~\mathrm{s.t.}~~~~~~~
\eqref{altitude cons}~\text{and}~\eqref{beamwidth cons}.\nonumber
\end{align}
It is worth noting that the constraints in \eqref{speed cons} can be approximately ignored in practice when the maximum UAV speed $V$ and/or the charging duration $T$ become sufficiently large. In Section IV, we present an efficient solution to problem (P1) based on the optimal solution to (P2).

\section{Optimal Solution to Problem (P2)}
In this section, we present the optimal solution to (P2). First, we show that solving (P2) is equivalent to solving the following problem (P3) that maximizes the average energy received by the two ERs.
\vspace{-0.5em}
\begin{align}
\text{(P3)}:\max_{\{x(t),H(t),\Theta(t)\}}~& \frac{1}{2} \sum_{k \in \mathcal{K}} \int_{0}^T {Q}_k(x(t),H(t),\Theta(t)) \mathrm{d} t \nonumber\\
\mathrm{s.t.}~~~~~
& \eqref{altitude cons}~\text{and}~\eqref{beamwidth cons}.\nonumber
\end{align}

\begin{lemma}\label{lemma1}
There exists an optimal solution to (P3) with the following structure:
\vspace{-0.5em}
\begin{align}\label{A}
&x(t)=-\bar{x}^*,\forall t \in (0,T/2],~x(t)=\bar{x}^*,\forall t \in (T/2,T],\nonumber\\
&H(t)=\bar{H}^*,\Theta(t)=\bar{\Theta}^*,\forall t \in \mathcal{T},
\end{align}
where $\bar x^*$, $\bar H^*$, and $\bar\Theta^*$ are variables to be decided later, such that the UAV adopts fixed beamwidth $\bar\Theta^*$ over the whole charging period, and hovers at two symmetric locations $(-\bar x^*,0,\bar H^*)$ and $(\bar x^*,0,\bar H^*)$ during each half of the period. The solution in \eqref{A} is also optimal for (P2).
\end{lemma}
\begin{IEEEproof}
The first part of this lemma follows directly due to the symmetric locations for the two ERs. In this case, by substituting the symmetric trajectory solution to (P2) and (P3), it is evident that they achieve the same objective value. Note that the optimal value achieved by the solution in \eqref{A} serves as an upper bound of that by (P2), and therefore, this solution is also optimal for (P2). Therefore, this lemma is proved.
\end{IEEEproof}
Based on Lemma \ref{lemma1}, we only need to get the optimal solution to (P3) by finding the optimal $\bar x^*$, $\bar H^*$, and $\bar \Theta^*$ in \eqref{A}, and accordingly obtain the optimal solution to (P2).

Now, we focus on solving (P3). Based on Lemma \ref{lemma1}, finding the optimal $\bar x^*$, $\bar H^*$, and $\bar \Theta^*$ corresponds to solving the following optimization problem over $\bar x$, $\bar H$, and $\bar \Theta$.
\vspace{-0.5em}
\begin{align}\label{subproblems}
&\max_{\bar x,\bar H,\bar \Theta}~ \frac{T}{2} \sum_{k \in \mathcal{K}}  {Q}_k(\bar x,\bar H,\bar \Theta) \\
&~~~\mathrm{s.t.}~~
 H_{\text{min}} \le \bar H \le H_{\text{max}},~\Theta_{\text{min}} \le \bar \Theta \le \Theta_{\text{max}}.\nonumber
\end{align}
It is evident that the optimal solution of $\bar x$ to problem \eqref{subproblems} must lie within the region $[-\frac{D}{2},\frac{D}{2}]$, since otherwise, the UAV can always move its horizontal location into this region to improve the objective value of problem \eqref{subproblems}. Also note that under any given $\bar H$ and $\bar \Theta$, $\bar x$ and $-\bar x$ achieve the same objective value of problem \eqref{subproblems}. As a result, in the following, we only need to focus on the solution with $\bar x \in [0,\frac{D}{2}]$ for problem \eqref{subproblems}.

Problem \eqref{subproblems} is still non-convex and thus difficult to be solved directly. To tackle this challenge, we solve problem \eqref{subproblems} by first considering two cases with $\tan \bar\Theta \ge \max_{k \in \mathcal{K}} |\bar x-\hat{x}_k|/\bar H$ and $\tan \bar\Theta <\max_{k \in \mathcal{K}} |\bar x-\hat{x}_k|/\bar H$, and then comparing them to obtain the optimal solution. Note that the two cases correspond to that the UAV can transfer energy to the two ERs simultaneously or can only transfer to one of the two ERs at each time instant, respectively.

\subsubsection{Case with $\tan \bar \Theta < \max_{k \in \mathcal{K}} |\bar x-\hat{x}_k|/\bar H$}
Since we only focus on the solution with $\bar x \in [0,\frac{D}{2}]$, the UAV only serves ER $2$ in this case. Accordingly, problem \eqref{subproblems} becomes
\vspace{-0.5em}
\begin{align}\label{case1 pro}
\max_{\bar x,\bar H,\bar \Theta}~& \frac{\beta_0G_0P}{2\bar \Theta^2[(\bar x-\hat{x}_2)^2+\bar H^2]}  \\
\mathrm{s.t.}~~~
& H_{\text{min}} \le \bar H \le H_{\text{max}},~\Theta_{\text{min}} \le \bar \Theta \le \Theta_{\text{max}},~0 \le \bar x \le \frac{D}{2}.\nonumber
\end{align}
Note that the objective function in \eqref{case1 pro} is monotonically decreasing with respect to $\bar H \in [H_{\text{min}},H_{\text{max}}]$ and $\bar \Theta \in [\Theta_{\text{min}},\Theta_{\text{max}}]$, and is monotonically increasing with respect to $\bar x \in [0,\frac{D}{2}]$. Therefore, the optimal solution to problem \eqref{case1 pro} is given as
\vspace{-1em}
\begin{align}\label{solu1}
\bar x^*_{1}=\frac{D}{2},~\bar H^*_{1}=H_{\text{min}},~\bar \Theta^*_{1}=\Theta_{\text{min}}.
\end{align}
Accordingly, the optimal objective value of problem \eqref{subproblems} in this case is
\vspace{-1em}
\begin{align}\label{opti1}
v_{1}=\frac{\beta_0G_0P}{2\Theta^2_{\text{min}}H^2_{\text{min}}}.
\end{align}

\subsubsection{Case with $\tan \bar \Theta \ge \max_{k \in \mathcal{K}} |\bar x-\hat{x}_k|/\bar H$}
In this case, problem \eqref{subproblems} is re-expressed as
\vspace{-0.5em}
\begin{align}\label{case2 pro}
\max_{\bar x,\bar H,\bar \Theta}~& \frac{1}{2} \sum_{k \in \mathcal{K}}  {Q}_k(\bar x,\bar H,\bar \Theta)  \\
\mathrm{s.t.}~~~
& H_{\text{min}} \le \bar H \le H_{\text{max}},~\Theta_{\text{min}} \le \bar \Theta \le \Theta_{\text{max}}\nonumber\\
& \tan \bar \Theta \ge (\bar x+D/2)/\bar H,~ 0 \le \bar x \le \frac{D}{2}.\nonumber
\end{align}
Notice that as $\tan \Theta_\text{max} \ge D/(2H_\text{max})$ is assumed, problem \eqref{case2 pro} is always feasible. Then we have the following lemma.

\begin{lemma}\label{lemma2}
At the optimal solution to problem \eqref{case2 pro}, it must hold that $\bar H\text{tan}\bar \Theta=\bar x+\frac{D}{2}$.
\end{lemma}
\begin{IEEEproof}
We have $\bar H\text{tan}\bar \Theta \ge \bar x+\frac{D}{2}$ to ensure that the two ERs are covered by the UAV simultaneously. Suppose $\bar H\text{tan}\bar \Theta > \bar x+\frac{D}{2}$ holds at the optimality. Then the UAV can always reduce its altitude or half-power beamwidth to improve the objective value in problem \eqref{case2 pro}. As a result, the presumption $\bar H\text{tan}\bar \Theta = \bar x+\frac{D}{2}$ holds. Therefore, this lemma is proved.
\end{IEEEproof}

Based on Lemma \ref{lemma2}, it follows that the UAV should set the beamwidth as the minimum one such that its transmitted signal beam can exactly cover the two ERs. Accordingly, we can reformulate problem \eqref{case2 pro} as
\vspace{-0.5em}
\begin{align}
\label{casetwo pro}
 \max_{\bar x,\bar \Theta} ~& \psi(\bar x,\bar \Theta) \triangleq \frac{\beta_0G_0P}{2\bar \Theta^2}\bigg[ \frac{1}{(\bar x-\frac{D}{2})^2+\frac{(\bar x+\frac{D}{2})^2}{\tan^2\bar \Theta}}+\frac{\sin^2\bar \Theta}{(\bar x+\frac{D}{2})^2} \bigg],\nonumber\\
\text{s.t.}~&~~~~~0 \le \bar x \le \frac{D}{2},~\Theta_{\text{min}} \le \bar \Theta \le \Theta_{\text{max}}\nonumber\\
&~~~~~H_{\text{min}} \le \frac{\bar x+\frac{D}{2}}{\tan\bar \Theta} \le H_{\text{max}}.
\end{align}

Problem \eqref{casetwo pro} is still non-convex due to the non-concavity of its objective function. To tackle this problem, in the following we first solve for $\bar x$ under given $\bar \Theta$, and then adopt a one-dimensional (1D) line search over $\bar\Theta \in [\check \Theta, \hat{\Theta}]$ to find the optimal solution of $\bar \Theta$, where $\check \Theta= \text{max}(\Theta_{\text{min}},\text{arctan}(\frac{D}{2H_{\text{min}}}))$ and $\hat{\Theta}=\text{min}(\Theta_{\text{max}},\text{arctan}(\frac{D}{H_{\text{max}}}))$ correspond to the lower and upper bounds of $\bar \Theta$.

Under any given $\bar \Theta$, problem \eqref{casetwo pro} can be expressed as
\vspace{-0.5em}
\begin{align}
\label{fix theta pro}
 \max_{\bar x} ~& \psi(\bar x,\bar \Theta) \nonumber\\
\text{s.t.}~~&\underline \alpha_{\bar \Theta} \le \bar x \le \overline \alpha_{\bar \Theta},
\end{align}
where $\underline \alpha_{\bar \Theta}=\text{max}(0,H_{\text{min}}\text{tan}\bar \Theta-\frac{D}{2})$ and $\overline \alpha_{\bar \Theta}=\text{min}(\frac{D}{2},H_{\text{max}}\text{tan}\bar \Theta-\frac{D}{2})$. Problem \eqref{fix theta pro} can be solved by examining the first-order derivative of $\psi(\bar x,\bar \Theta)$ with respect to $\bar x$ as follows.
\vspace{-0.5em}
\begin{align}
\label{x first order}
\frac{\partial \psi(\bar x,\bar \Theta)}{\partial \bar x}=
-\frac{\beta_0G_0P}{\bar \Theta^2} \bigg[ \frac{\bar x-\frac{D}{2}+\frac{\bar x+\frac{D}{2}}{\tan^2\bar \Theta}}{[(\bar x-\frac{D}{2})^2+\frac{(\bar x+\frac{D}{2})^2}{\tan^2\bar \Theta}]^2}+\frac{\sin^2\bar \Theta}{(x+\frac{D}{2})^3} \bigg].
\end{align}
Notice that $\frac{\partial \psi(\bar x,\bar \Theta)}{\partial \bar x} = 0$ corresponds a polynomial equation of degree 4. Therefore, this equation has a total of 4 solutions at maximum. Suppose that $\tilde \alpha_{1,\bar \Theta}, \dots, \tilde \alpha_{m,\bar \Theta}$ denote the solutions of $\bar x$ to this equation within the region of $[\underline \alpha_{\bar \Theta},\overline \alpha_{\bar \Theta}]$, where $m \le 4$ in general. Then we have the following lemma.
\begin{lemma}\label{lemma3}
The optimal solution to problem \eqref{fix theta pro} is given as
\vspace{-1em}
\begin{align}
x^{(\bar \Theta)}=\begin{cases} \underline \alpha_{\bar \Theta},~&\text{if}~ \bar \Theta \in [\check{\Theta}, \pi/4]\\\mathop{\arg\max}_{\bar x \in \mathcal{X}_{\bar \Theta}} \psi(\bar x,\bar \Theta),~&\text{if}~ \bar \Theta \in (\pi/4, \hat{\Theta}],\end{cases}
\end{align}
where $\mathcal{X}_{\bar \Theta}= \{\tilde \alpha_{1,\bar \Theta},\dots,\tilde \alpha_{m,\bar \Theta},\underline \alpha_{\bar \Theta},\overline \alpha_{\bar \Theta}\}$.
\end{lemma}
\begin{IEEEproof}
When $\bar \Theta \in [\check{\Theta}, \pi/4]$, it is evident that $\frac{\partial \psi(\bar x,\bar \Theta)}{\partial \bar x}<0,\forall \bar x \in [\underline \alpha_{\bar \Theta}, \overline \alpha_{\bar \Theta}],$ holds, and thus $\psi(\bar x,\bar \Theta)$ is monotonically decreasing over $\bar x \in [\underline \alpha_{\bar \Theta}, \overline \alpha_{\bar \Theta}]$. As a result, we have $x^{(\bar \Theta)}= \underline \alpha_{\bar \Theta},~\text{if}~ \bar \Theta \in [\check{\Theta}, \pi/4]$. On the other hand, when $\bar \Theta \in (\pi/4,\hat{\Theta}]$, the optimal solution to problem \eqref{fix theta pro} is included in $\mathcal{X}_{\bar \Theta}$ due to the  continuousness of $\psi(\bar x,\bar \Theta)$ in the region of $[\underline \alpha_{\bar \Theta},\overline \alpha_{\bar \Theta}]$. By comparing all the elements in $\mathcal{X}_{\bar \Theta}$, we can obtain the solution that achieves a larger objective value for problem \eqref{fix theta pro}, i.e., $x^{(\bar \Theta)}=\mathop{\arg\max}_{\bar x \in \mathcal{X}_{\bar \Theta}}~ \psi(\bar x,\bar \Theta),~\text{if}~ \bar \Theta \in (\pi/4, \hat{\Theta}]$. Therefore, this lemma is proved.
\end{IEEEproof}

With the optimal solution of $\bar x$ to problem \eqref{casetwo pro} obtained under any given $\bar \Theta \in [\check{\Theta},\hat{\Theta}]$, together with the 1D line search over $\bar \Theta \in [\check{\Theta},\hat{\Theta}]$, we have the optimal solution of $\bar \Theta$ to problem \eqref{casetwo pro} as $\bar \Theta^*_{2}=\mathop{\arg\max}_{\bar \Theta \in [\check{\Theta},\hat{\Theta}]}~\psi(x^{(\bar \Theta)},\bar \Theta)$, and the corresponding solution of $\bar x$ and $\bar H$ as $\bar x^*_{2}=x^{(\bar \Theta^*_{2})},\bar H^*_{2}=\frac{x^{(\bar \Theta^*_{2})}+\frac{D}{2}}{\tan \bar \Theta^*_{2}}$. Therefore, the optimal value of problem \eqref{case2 pro} in this case is given as
\vspace{-0.5em}
\begin{align}\label{opti2}
v_{2}=\frac{\beta_0G_0P}{2(\bar \Theta^*_2)^2}  \bigg[ \frac{1}{(\bar x^*_2-\frac{D}{2})^2+(\bar H^*_2)^2}+\frac{1}{(\bar x^*_2+\frac{D}{2})^2+(\bar H^*_2)^2} \bigg].
\end{align}

\subsubsection{Optimal Solution to Problem \eqref{subproblems}}
Finally, by comparing the optimal values $v_1$ in \eqref{opti1} and $v_2$ in \eqref{opti2} for the above two cases, the optimal solution to problem \eqref{subproblems}, denoted by $\bar x^*$, $\bar H^*$, and $\bar \Theta^*$, is obtained as follows. If $v_1 \ge v_2$, then it follows that $\bar x^*=\bar x^*_1,~\bar H^*=\bar H^*_1,~\bar \Theta^*=\bar \Theta^*_1$; otherwise, we have $\bar x^*=\bar x^*_2,~\bar H^*=\bar H^*_2,~\bar \Theta^*=\bar \Theta^*_2$.

Based on the optimal solution to problem \eqref{subproblems} together with Lemma \ref{lemma1}, the optimal solution to (P3) and thus (P2), denoted by $\{x^{\star\star}(t)\}$, $\{H^{\star\star}(t)\}$, and $\{\Theta^{\star\star}(t)\}$, is given as
\vspace{-0.5em}
\begin{align}\label{opti solution}
&x^{\star\star}(t)=-\bar x^*,\forall t \in [0,T/2],~x^{\star\star}(t)=\bar x^*,\forall t \in (T/2,T],\nonumber\\
&H^{\star\star}(t)=\bar H^*,~\Theta^{\star\star}(t)=\bar \Theta^*,\forall t \in \mathcal{T}.
\end{align}

\begin{remark}
From the optimal solution in \eqref{opti solution}, it is evident that at the optimality, the UAV should adopt fixed beamwidth over the whole charging period, and hover at two symmetric locations with identical altitude during each half of the period. More specifically, if $\tan \bar \Theta^* <\max_{k \in \mathcal{K}}|\bar x^* - \hat x_k|/\bar H^*$ (i.e., the UAV only serves one ER at a time), then the UAV should hover above one ER at the lowest altitude with the minimum beamwidth (see \eqref{solu1} and \eqref{opti1}), so as to maximize the antenna gain and minimize the pathloss. Otherwise, if  $\tan \bar \Theta^* \ge \max_{k \in \mathcal{K}} |\bar x^* - \hat x_k|/\bar H^*$ (i.e., the UAV serves two ER simultaneously), then the beamwidth and the altitude should be properly designed to balance the tradeoff between the antenna gain and the pathloss. For convenience, we name the optimal trajectory solution to (P2) as the symmetric-location hovering solution.
\end{remark}

\section{Proposed Solution to Problem (P1)}
In this section, we consider (P1) with the maximum UAV speed constraints in \eqref{speed cons} considered. As (P1) is difficult to be solved optimally, we propose a heuristic hover-fly-hover trajectory based on the optimal solution to problem (P2) obtained in the previous section. Let $\{x^\star(t)\}$, $\{H^\star(t)\}$, and $\{\Theta^\star(t)\}$ denote the obtained solution to problem (P1), and $E^\star$ denote the common energy received by the two ERs over the whole charging period. In the following, we consider the two cases when there are one and two hovering locations with $\bar x^*=0$ and $\bar x^*>0$ for (P2), respectively.

When $\bar x^*=0$, the UAV only needs to hover at one single location, and as a result, the optimal solution to (P2) is also optimal to (P1), i.e.,
\vspace{-0.5em}
\begin{align}
x^\star(t)=0,~H^\star(t)=\bar H^*,~\Theta^\star(t)=\bar \Theta^*,\forall t \in \mathcal{T}.
\end{align}
The common energy received by the two ERs is
\vspace{-0.5em}
\begin{align}
E^\star=\frac{\beta_0G_0PT}{(\bar \Theta^*)^2[(\frac{D}{2})^2+(\bar H^*)^2]}.
\end{align}
Next, we consider the case with $\bar x^*>0$, in which the UAV needs to hover at two symmetric locations at the optimal solution to (P2). In this case, the optimal solution to (P2) is not feasible to (P1) due to the maximum speed constraints involved. To tackle this issue, we propose a hover-fly-hover trajectory design, in which the UAV first hovers at location $(-\bar x^*,0,\bar H^*)$ for a certain duration of $T_\text{hover}$, then flies from $(-\bar x^*,0,\bar H^*)$ to $(\bar x^*,0,\bar H^*)$ with fixed altitude $\bar H^*$ and maximum speed $V$, and finally hovers at location $(\bar x^*,0,\bar H^*)$ for the remaining time with same hovering duration $T_\text{hover}$. During flying, the UAV adopts the fixed altitude and the maximum speed for minimizing the flying time.

In this case, the travel distance between the two locations is $2\bar x^*$, and the flying duration is $T_{\text{fly}}=\frac{2\bar x^*}{V}$. Consequently, the hovering durations at the two locations are equally allocated as $T_{\text{hover}}=\frac{T}{2}-\frac{\bar x^*}{V}$. In this case, the corresponding UAV trajectory and altitude are respectively expressed as\footnote{Note that in this paper we only consider the case with $T \ge T_{\text{fly}}$, which corresponds to that $T$ is sufficiently long for efficiently charging the two ERs in practice.}
\vspace{-0.5em}
\begin{align}\label{tra}
x^\star(t)=\begin{cases}-\bar x^*,&\forall t \in [0,\frac{T}{2}-\frac{\bar x^*}{V}]\\Vt-VT/2,&\forall t \in (\frac{T}{2}-\frac{\bar x^*}{V},\frac{T}{2}+\frac{\bar x^*}{V})\\\bar x^*,&\forall t \in [\frac{T}{2}+\frac{\bar x^*}{V},T],\end{cases}
\end{align}
\vspace{-1em}
\begin{align}\label{altitude}
H^\star(t)=\bar H^*,\forall t \in \mathcal{T}.
\end{align}

Next, we determine the beamwidth $\{\Theta^\star(t)\}$. During hovering, it follows that
\vspace{-0.5em}
\begin{align}\label{B}
\Theta^\star(t) = \bar\Theta^*, \forall t\in [0,\frac{T}{2}-\frac{\bar x^*}{V}] \cup [\frac{T}{2}+\frac{\bar x^*}{V}, T].
\end{align}
During flying, notice that under given altitude $H^\star(t)$ and location $x^\star(t)$ at time $t \in (\frac{T}{2}-\frac{\bar x^*}{V},\frac{T}{2}+\frac{\bar x^*}{V})$, the UAV prefer to set the beamwidth as narrow as possible to maximize the antenna gain, provided that the corresponding ERs are covered properly. Let $\Theta^\star_1(t)=\arctan \frac{\min_{k \in \mathcal{K}}|x^\star(t)-\hat{x}_k|}{\bar H^*}$ denote the minimum beamwidth when the UAV only serves one ER, and $\Theta^\star_2(t)=\arctan \frac{\max_{k \in \mathcal{K}}|x^\star(t)-\hat{x}_k|}{\bar H^*}$ denote the minimum beamwidth when the UAV serves the two ERs simultaneously. Then we choose the one that achieves a higher received power by the two ERs, i.e.,
\vspace{-0.5em}
\begin{align}\label{C}
\Theta^\star(t)=&\begin{cases}\Theta^\star_1(t),&\text{if} ~\sum_{k \in \mathcal{K}}Q_k(x^\star(t),H^\star(t),\Theta^\star_1(t)) \ge\\
 &~~~\sum_{k \in \mathcal{K}}Q_k(x^\star(t),H^\star(t),\Theta^\star_2(t))\\
 \Theta^\star_2(t),&\text{if}~\sum_{k \in \mathcal{K}}Q_k(x^\star(t),H^\star(t),\Theta^\star_1(t))<\\
 & ~~~\sum_{k \in \mathcal{K}}Q_k(x^\star(t),H^\star(t),\Theta^\star_2(t)),\nonumber)\end{cases}\nonumber\\
  &~~~~~~~~~~~~~~~~~~\forall t \in (\frac{T}{2}-\frac{\bar x^*}{V},\frac{T}{2}+\frac{\bar x^*}{V}).
\end{align}
By combing the solutions in \eqref{tra}, \eqref{altitude}, \eqref{B}, and \eqref{C}, the common energy received by the two ERs is
\vspace{-0.5em}
\begin{align}
E^\star=\frac{1}{2}\int_{0}^{T} \sum_{k \in \mathcal{K}}  {Q}_k(x^\star(t),H^*,\Theta^\star(t))\text{d}t.
\end{align}

\begin{remark}\label{remark:1}
Note that when the charging duration $T$ becomes large, our proposed hover-fly-hover trajectory design is asymptotically optimal for (P1). This is due to the fact that in this case, the flying time $T_{\text{fly}}$ becomes negligible, and thus the maximum common energy achieved by the hover-fly-hover trajectory design approaches the optimal objective value of (P1).
\end{remark}

\section{Numerical Results}
In this section, we present numerical results to validate the performance of our proposed design, as compared to the following two schemes.
\setcounter{subsubsection}{0}

\subsubsection{Convention UAV-enabled WPT design with omni-directional antenna \cite{XuZengZhang20171}}
The UAV is equipped with an omni-directional antenna that has a unit antenna gain. It is evident that the UAV should fly at the minimum altitude with $H(t) = H_{\text{min}},\forall t\in \mathcal{T}$. In this case, the trajectory optimization for maximizing the common energy received by the two ERs has been investigated in \cite{XuZengZhang20171}, which shows that the optimal trajectory solution always follows a hover-fly-hover structure.

\subsubsection{Static hovering benchmark}
The UAV hovers at one fixed location $x^\text{sta}= 0$ over the whole charging period to balance the received energy at the two ERs, i.e., $x(t) = 0,\forall t\in\mathcal T$. The altitude and beamwidth of the UAV also remain unchanged with $H(t)=H^{\text{sta}},\Theta(t)=\Theta^{\text{sta}},\forall t \in \mathcal{T}$.
In the case with $\tan \Theta_{\text{min}} \ge D/(2H_{\text{min}})$, the UAV's transmitted signal beam can always cover the two ERs even with the minimum altitude and beamwidth adopted, and thus $H^{\text{sta}}=H_{\text{min}}$, and $\Theta^{\text{sta}}=\Theta_{\text{min}}$ apply. In the case with $\tan \Theta_{\text{min}} < D/(2H_{\text{min}})$, it must hold that $H^\text{sta}=\frac{D}{2\tan\Theta^\text{sta}}$ at the optimal solution; accordingly, we have $\Theta^{\text{sta}}=\text{max}(\Theta_{\text{min}},\text{arctan}(\frac{D}{2H_{\text{max}}}))$ and $H^{\text{sta}}=\text{min}(H_{\text{max}},\frac{D}{2\text{tan}\Theta_{\text{min}}})$.

In the simulation, we set $\beta_0=-30~\text{dB}$, $H_{\text{min}}=10 ~\text{m}$, $H_{\text{max}}=30 ~\text{m}$, $\Theta_{\text{min}}=\pi/6$, $\Theta_{\text{max}}=\pi/2$, $P=40~\text{dBm}$, $V=5 ~\text{m/s}$, unless specified otherwise.

\begin{figure}
\centering
 \epsfxsize=1\linewidth
    \includegraphics[width=6.5cm]{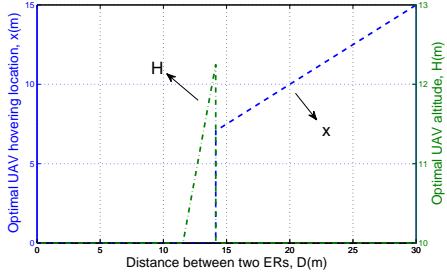}
\caption{The optimal UAV hovering location $\bar x^*$ and the optimal UAV altitude $\bar H^*$ versus the distance $D$ between the two ERs.}\vspace{-1em}
\end{figure}

Fig. 1 shows the optimal UAV hovering location $\bar x^*$ and the optimal UAV altitude $\bar H^*$ versus the distance $D$ between the two ERs. It is observed that when $D$ is small (e.g., $D \le 12~\text{m}$), the UAV prefers to hover above the middle point of the two ERs (i.e., $\bar x^* = 0$) with the lowest altitude and the minimum beamwidth to serve them simultaneously. It is also observed that when $D$ is large (e.g., $D \ge 15~\text{m}$), the UAV prefers to hover exactly above each ER for serving them individually with the lowest altitude and the minimum beamwidth. When $D$ is moderate (e.g., $12~\text{m} \le D \le 14 ~\text{m}$), the UAV is observed to adjust the hovering location and altitude (together with the beamwidth as well) to balance the tradeoff between the antenna gain and pathloss for WPT performance optimization.

\begin{figure}
\centering
 \epsfxsize=1\linewidth
    \includegraphics[width=6.5cm]{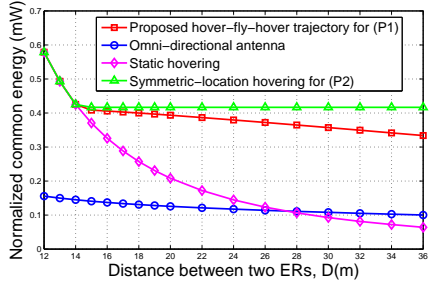}
\caption{Normalized common energy received by two ERs versus the distance $D$.}\vspace{-2em}
\end{figure}

Fig. 2 shows the common energy received by the two ERs (normalized by the charging duration $T$) versus the distance $D$, where $T=20~\text{s}$. It is observed that when $D$ is small (e.g., $D \le 14~\text{m}$), our proposed hover-fly-hover trajectory and the static-hovering benchamark for (P1) achieve the same common received energy value as the upper bound achieved by the symmetric-location hovering for (P2). This shows that in this case, it is optimal for the UAV to hover above the middle point between the two ERs for efficient WPT, as shown in Fig. 1. It is also observed that when $D$ becomes large (e.g., $D \ge 15~\text{m}$), our proposed design outperforms both the conventional design with omni-directional antenna and the static hovering benchmark. This validates the benefit of jointly exploiting the UAV mobility and using directional antenna with altitude and beamwidth control.

\begin{figure}
\centering
 \epsfxsize=1\linewidth
    \includegraphics[width=6.5cm]{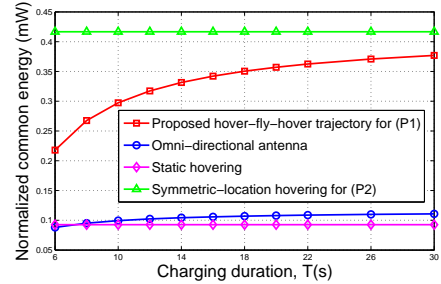}
\caption{Normalized common energy received by two ERs versus the charging duration $T$.}\vspace{-2em}
\end{figure}

Fig. 3 shows the normalized common energy received by the two ERs versus the charging duration $T$, where $D=30~\text{m}$. It is observed that our proposed hover-fly-hover trajectory design outperforms the conventional design with omni-directional antenna and the static hovering benchmark, and the gain becomes more substantial when $T$ becomes larger. It is also observed that our proposed design approaches the performance upper bound by the symmetric-location hovering for (P2) as $T$ increases. This is consistent with Remark \ref{remark:1}.

\section{Conclusion}

This paper studied a UAV-enabled WPT system, in which a UAV equipped with a directional antenna is dispatched to deliver energy to charge two ERs on the ground. We aimed to maximize the common energy received by the two ERs by jointly optimizing the UAV's trajectory, altitude, and beamwidth over one particular finite charging period, subject to the maximum UAV speed constraints, as well as the maximum/minimum altitude and beamwidth constraints at the UAV. Efficient algorithms were proposed to solve the common energy maximization problem. Numerical results were provided to validate the benefit of directional antenna for UAV-enabled WPT, as compared to other benchmark schemes.

\end{document}